Temperature Effect on the Interactions between Oil Droplet and Kerogen Surface at Reservoir Temperatures

Zelong Zhang,*,† Adrienne Stephens,† and Jianwei Wang†,§

†Department of Geology and Geophysics, Louisiana State University, Baton Rouge, LA 70803, United States

§Center for Computation and Technology, Louisiana State University, Baton Rouge, LA 70803, United States

*Corresponding to: zhangzelong@protonmail.com

**Abstract**

Understanding the thermodynamics of the interfacial interactions between oil and kerogen is imperative for recovering hydrocarbon in tight reservoirs, especially in unconventional shale that retains abundant hydrocarbon in kerogen nanopores. The temperature effect on the interactions of light oil with a type II kerogen in water was investigated using molecular dynamics simulation. Non-polar and polar light oil droplets were modeled with clusters of 30 octane molecules and 30 octanethiol molecules, respectively. Kerogen was modeled with a molecular fragment from a type II kerogen. The free energy calculations were performed at constant volume and temperature with umbrella sampling at temperatures in the range of 300–500 K (27–227 °C, 80–440 °F), comparable to the reservoir conditions of common shale plays. The result shows that the free energy of desorption of an oil droplet scales linearly with temperature. For oil droplets, the desorption free energy cannot be quantitatively scaled up from that of a single oil molecule. Additionally, the free energy of desorption exhibited a strong temperature dependence, suggesting a significant entropic contribution to the free energy. The contact angle of oil droplets was estimated by the morphologies of the oil cluster in contact with the kerogen surface, identified at the lowest free energy point in the free energy profile. The cosine of the contact angle is linearly correlated with the free energy of the desorption. This study provides a thermodynamic basis and molecular details on how temperature affects the oil interactions with kerogen, providing a valuable insight to strategy for improving unconventional oil recovery.

## Introduction

Petroleum is a major energy source and therefore crude oil is a strategic resource[1–3] Current technology can yield up to 30% –60% of the original oil in place, leaving up to 70% crude oil in a reservoir.[4,5] For unconventional tight reservoirs, the fraction of oil recovered is much lower. A large fraction of uncovered oil remains in porous media of minerals and organic materials in the rock, especially for tight formations where most uncovered oil is adsorbed at the surfaces of the nanopore network of materials such as kerogen.[6–9] The intermolecular interactions at the surfaces play an important role in confining the oil-containing fluid and keeping the oil molecules from being recoverable.[10,11] Thus, thermodynamics of the fluid-surface interactions is critical for understanding the behavior of the fluid and the impact of physical conditions and fluid composition, which are essential for developing strategies for unconventional shale oil.[12–14]

Oil recovery technologies have applied heat to enhance the recovery process by steam injection, hot water flooding, and in-situ combustion.[15] The thermal stimulations induced temperature effect on the properties of the fluid at the interfaces may play a critical role in the unconventional oil recovery. These techniques are originally designed for recovering conventional hydrocarbon, especially heavy oil. Recent studies based on reservoir modeling and laboratory observation show that thermal stimulations can also be economically viable to improve oil recovery from unconventional shale.[16,17] In these models, fluid properties, such as viscosity, density, and compressibility, were based on the experimental values.[18,19] The impact of fluid-surface interactions on the fluid properties at the molecular scale are not explicitly included in the models to describe the fluid transport. Such a treatment of the fluid properties could significantly affect the accuracy of the models by ignoring the interfacial interactions, especially for the fluid in abundant nanopores of kerogen-rich shale, where the fraction of the fluid affected by the surfaces is increasingly high as the pore size decreases (ref your paper here). The interfacial interactions within the nanopores require the atomistic scale treatment. Indeed, Wang et al., 2015 evaluated the temperature effect on the density distribution of octane in kerogen slit (graphene surface) at temperature ranging from 333 to 393 K.[20] They observed an up to 2% reduction in the number of adsorbed octane molecules at elevated temperatures. Recently, Yang et al., 2020 examined the temperature effect on the adsorption of hydrocarbon mixtures at the surfaces of a kerogen slit (porous functionalized surface) at temperatures ranging from 280 to 400 K. They observed a linear reduction in the number of oil molecules confined in the slit as a function of temperature.[21] Both of these findings are based on the phenomena. Decreased adsorptions of oil molecules were observed at elevated temperatures. However, the thermodynamics of the interfacial fluid, mechanisms of the fluid desorption, and energetic factors contributed to the increased recovery remain unclear.

Desorption of oil molecules from the surfaces in shale is a thermodynamic process. Data on the free energy of the process is essential for the understanding of both the equilibrium and kinetics. How does temperature affect the free energy of desorption? What are the enthalpy and entropic contributions to the free energy? Answers to these questions will provide a fundamental basis to the thermal stimulation techniques. It has been shown that the entropic contribution to the desorption free energy is important for the understanding of the desorption kinetics for organic molecules on some oxide and metal surfaces.[22] The entropy of adsorption at a number of metal oxide surfaces can be predicted from their gas phase entropy for alkane and alcohol molecules. For zeolite framework structures, the entropy of adsorption can be predicted from only a single material descriptor, the occupiable volume for hydrocarbon molecules.[22,23] Giving the similarity on the molecular adsorptions at solid surfaces, we hypothesized that the entropic contribution to the free energy of oil adsorption at kerogen surface is significant in controlling the free energy of the desorption of oil molecules from the kerogen surface.

In the present study, the temperature effect on the free energy of interactions of light oil molecules with shale kerogen surfaces was systematically investigated at temperatures from 300 to 500 K. The enthalpy and entropic contributions to the free energy were estimated. The effect of molecular polarity (polar vs non-polar oil) and molecular clustering (oil droplet vs single oil molecule) on the desorption free energy were taken into consideration. In addition, the correlation between the free energy and the contact angles of oil droplet was explored to understand the wettability of oil droplets at kerogen surface. Quantifying the relationships between free energy, temperature, and contact angle provides a thermodynamic basis to understand the interface wettability and thermal stimulation in terms of recovering hydrocarbon fluid confined in nanoporous shale.

## Computational Methods

**Molecular Models.** Two types of interfacial system were investigated: a single oil molecule and an oil droplet (modeled by a cluster of 30 molecules) on a water-wetted kerogen surface (Figure 1). The size of the simulation box, number of water molecules, and kerogen slab thickness are tabulated in Table 1. The input structures of water, oil, oil cluster, and kerogen surface were taken from our previous study for modeling of adsorption at kerogen surface.[24] Hydrocarbons with eight carbons are one of the most abundant compounds in crude tight oil,[25] especially in the Bakken formation.[26] Therefore, n-octane ($C_8H_{18}$) was selected to model the light oil compound. Given that the sulfur content is abundant in crude oil,[27,28] octanethiol ($C_8H_{17}SH$) was used as a model for sulfur-containing oil and a polar oil molecules to evaluate the effect of molecular polarity on the desorption. The oil droplets were prepared using 30 molecules of octane and octanethiol to represent the non-polar and polar oil droplets, respectively. A type II kerogen slab was built with 511 kerogen fragment molecules ($C_{22}H_{13}ON$), which is adapted from a thermally mature type II kerogen molecule.[29] The assembled kerogen slab has an H/C ratio of 0.59 and an O/C ratio of 0.05, which is consistent with the definition of type II kerogen. The density of this kerogen slab is 1.15 g/cm$^3$, close to the experimental data: 1.18 – 1.89 g/cm$^3$.[30–33] Detailed procedures on how to prepare the kerogen slab surface were reported in our previous publication.[24]

Because the kerogen model surface is structurally flexible and is sensitive to the temperature change. Therefore, increasing the temperature can roughen the model surface. High roughness can lead to a technical difficulty in locating the surface plane and constraining the distance between the molecules and surface in umbrella sampling, which creates a challenge to reliably calculate the free energy profile of the desorption. Thus, the kerogen surfaces were rigid by fixing the position of all kerogen atoms in all directions (X, Y, and Z). However, the rigid surfaces create an artificial effect on the dynamics of the interfacial interactions by altering the momentum transfer across the interface, which can induce a systematic error in the calculated desorption free energy. To estimate the error caused from using rigid kerogen surface, we compared the free energy of desorption of a single polar oil molecule calculated without fixing the kerogen surface at relatively lower temperatures, i.e., 200K to 400K, with the values calculated with fixed kerogen surface at the same temperatures. As shown in Figure S1, the free energies calculated at the fixed kerogen surface are consistently overestimated by a constant, ~3-6 kJ/mol, independent of temperature within the temperature range. The result suggests that the rigid surface can be used calculate the dependence of the free energy of adsorption as a function of temperature with the absolute free energy shifted by an average of +4.5 kJ/mol.

There is a collection of molecular models have been used for water. Although there is not a single water model that can produce all water properties accurately,[34,35], most water models can produce results qualitatively well.[34] Nonetheless, we compared several common water models by replicating the calculation of free energy profile of the single polar oil molecule desorption at the wet kerogen surface. Common water models we tested include simple point charge series (SPC, SPC/F, and SPC/E) and transferable intermolecular potential series (TIP3P, TIP4P-Ew, and TIP5P-E). The calculated results and performance details are compared in Figure S2 and Table S1. Different water models produced noticeable differences in the free energy. The highest desorption energy was produced by TIP4P-Ew model and the lowest desorption energy by SPC/F, which is attributed to the large dipole moment of water model induced by the flexible structure.[36] However, they all share qualitatively similar profiles., This study used SPC/F water model, considering its simplicity and the consequent improvement in computational efficiency.[37–40] The SPC/F model is also adopted in CLAYFF force field,[41] which has been extensively used to describe aqueous solution interactions with mineral surfaces.[42–44] The parameters of SPC/F potential used in this study are listed as supporting text 1. It is anticipated that using different water models will not alter how the free energy responds to the temperature change, although the absolute values in free energy may change.

The OPLS-AA force field was applied to describe the organic molecules including octane, octanethiol, and kerogen.[45] Their molecular geometries and dipole moments described by this force field are in agreement with results from density function theory calculation, as shown in Table S2. The interactions between the oil molecules, kerogen, and water were calculated using the combination rule based geometric averages. All these force field potentials have produced reasonable results as demonstrated by our previous study.[24]

**Molecular Dynamics Simulation and Free Energy Calculation.** Molecular dynamics (MD) simulations were carried out with GROMACS.[46–52] The free energy profile of oil desorption was computed using the umbrella sampling. All MD simulations were performed in canonical ensembles (NVT) with the following settings: periodic boundary conditions, time step of 1.0 fs, fast smooth particle-mesh Ewald (SPME) for electrostatic interaction with interpolation order of 4, relative strength of the Ewald-shifted direct potential $10^{-5}$, 0.12 nm fourierspacing, Verlet cutoff-scheme, and a Nosé–Hoover extended ensemble for temperature coupling every 0.41 ps. Five different temperatures were investigated, including 300 K, 350 K, 400 K, 450 K, and 500 K. The temperature range is based on the reported data of common shale plays of temperature from 305 K to 436 K.[53] As shown in Figure 1, slab models were used, a single molecule and molecular cluster are fully immersed in water molecules, with a sufficient gap between the other kerogen surface and water (Figure 1). Despite the presence of vacuum, oil molecule and droplets in all simulations were constantly being fully submerged in water. No phase changes have been observed during these simulations.

All systems were initially equilibrated prior to umbrella sampling. The center of mass of the oil molecules was fixed at a distance along the direction of the reaction coordinate. Each simulation was run ~100 ps to allow the system to relax and reach equilibrium. The setting of umbrella sampling simulation for each system is tabulated in Table 1. Upon the completion of umbrella sampling simulations, weighted histogram analysis method (WHAM) was carried out to compute the free energy profile (i.e. potential of mean force) and error analysis. The errors were estimated using Bayesian bootstrapping. Each bootstrapping used 10 bootstraps. The free energy calculated from the NVT ensemble is Helmholtz free energy ($\Delta A$). Since a gap was maintained between water and one side of the kerogen surfaces, pressure is assumed to be constant for the simulations. In this context, free energy or Gibbs free energy is used throughout this study.

**Data Visualization and Analysis.** The simulation data were visualized by VMD.[54] VMD was also used to analyze the surface area of interfaces and the quantity of adsorbed oil atoms in the first layer on surface through Tcl/tk scripting (Supporting Text 2). The surface area was measured using solvent-accessible surface area (SASA) algorithm with a probe radius. Oil atoms within 0.36 nm distance of the kerogen surface were counted as the first layer adsorbates, considering the maximum range of non-bonded interactions is approximate to 0.36 nm in these simulations. The contact angles of oil droplets were measured by Fiji ImageJ (version 1.52p, open source)[55] using contact angle plugin developed by Marco Brugnara. Circle best-fit algorithm was applied. The reported surface areas and contact angles were averaged from the estimations of trajectory snapshots from simulations at the free energy minimum or at the adsorbed state. The contact angle measurements were listed in the Supporting Information as Table S3.

## Results and Discussion

### Effect of Temperature on the Free Energy of Desorption

The calculated free energy of desorption of oil droplets are shown in Figure 2 and Table 2. As temperature increases, the desorption free energy decreases in an approximately linear fashion for both the polar and non-polar oil droplets. The result is consistent with a previous study suggesting that entropies of alkane adsorption on acidic zeolites are insensitive over a moderate temperature range up to 450 K.[56] If there is a linear relationship between free energy and temperature within the temperature range investigated, both the enthalpy and entropy are constant, and the following equation can then be used:

$$\Delta G(T) = \Delta H - T\Delta S \quad (1)$$

where $\Delta G$ (kJ/mol) is the change of free energy of desorption, $\Delta H$ (kJ/mol) is the enthalpy, $T$ is the temperature (K), and $\Delta S$ (J/mol/k) is the entropic contribution to free energy. Since the reported reservoir temperatures of common shale plays are between 305 K and 436 K[53] within a temperature range less than 150 K, equation (1) can then be used to calculate $\Delta H$ and $\Delta S$ by linear fitting.

As listed in table 2, the enthalpy of desorption $\Delta H$ is +14.3 (0.5) kJ/mol and +15.0 (1.3) kJ/mol per molecule for polar and non-polar droplets, respectively. The entropy of desorption $\Delta S$ is +22.7 (1.4) J/mol/K and +26.3 (3.2) J/mol/K per molecule for the polar and non-polar oil droplets, respectively. The positive enthalpy indicates that oil droplet desorption from wet kerogen is an endothermic reaction. The magnitude of $\Delta H$ is a quantitative measure of the strength of the binding between adsorbate (i.e., oil) and adsorbent (i.e., kerogen surface).[57 58] In comparison with the molecular adsorption of hydrocarbon molecules in zeolite framework structures,[23] the values (14.3, 15.0 kJ/mol) are at the lower end of adsorption enthalpy, suggesting weaker interactions between the oil droplets and kerogen surface with respect to the adsorption in zeolite framework. The comparable values (14.3, 15.0 kJ/mol) between polar and non-polar droplets suggest a similar strength of the binding with the kerogen surface. This is largely because kerogen surface exhibits both hydrophobic and hydrophilic molecular groups that have similar interactions with either octane or octanethiol droplets.

The positive $\Delta S$ suggests that the oil molecules gain entropy upon desorption, indicating the oil molecules are less constrained with increased randomness and disorder at desorbed state than at adsorbed state. For a short chain alkane with less than ten carbons, surface adsorption has no impact on the vibrational freedom.[22] Since the oil molecules, octane and octanethiol, have eight carbons, the entropy of desorption can be mostly attributed to the translational and the rotational freedoms. The small but statistically significant difference in $\Delta S$ between octane (26.3 J/mol/k) and octanethiol (22.7 J/mol/k) suggests that the octane droplet gains more entropy after desorption, largely because more hydrophobic nature of the molecule in water. It is interesting to note that the calculated $\Delta H$ and $\Delta S$ values for the oil droplets follow the trend of the relationship between enthalpy and entropy of molecular adsorption, which is established for alkanes in zeolite framework structures.[23] This result implies that the surface adsorption process of the oil droplets at kerogen surface behaves similarly to molecular adsorptions confined in the internal structure of zeolites.[23]

As Figure 2 and Table 2 suggested, increasing temperature reduces desorption free energy and promotes the release of oil molecules from the surface. Such understanding provides a fundamental basis for oil recovery technologies by applying thermal stimulations such as steam injection, hot water flooding, and in-situ combustion.[15] Because the $\Delta S$ gives the temperature effect on the desorption free energy, understanding of the entropy of desorption of oil molecules in water or brine at reservoir conditions may provide important clues to improve the effectiveness of the thermal stimulation in enhanced oil recovery. Since there is a strong correlation between the adsorption entropy of molecules at MgO surface and the entropy of the gas-phase molecule,[15] exploring the gas-phase entropy of oil molecules may provide valuable insight on thermal enhanced oil recovery.

**Effect of Molecular Cluster**

To understand how a molecular cluster affects the desorption free energy, the desorption of a single oil molecule in water at the kerogen surface was investigated for comparison. The free energy is plotted in Figure 3 and the enthalpy and entropy are listed in Table 2. The $\Delta H$ and $\Delta S$ were calculated by a linear fitting using equation (1). The enthalpy of desorption $\Delta H$ is 21.6 (1.5) kJ/mol and 20.6 (4.8) kJ/mol per molecule for polar and non-polar droplets, respectively. The entropy of desorption $\Delta S$ is 6.6 (3.7) J/mol/K and 8.2 (11.4) J/mol/K per molecule for the polar and non-polar oil droplets, respectively. The errors in entropy and enthalpy are relatively large because the small systems of a single oil molecule are more susceptible to statistical errors than the large systems of an oil droplet.[59] Considering the errors of the calculations, both $\Delta H$ and $\Delta S$ are statistically indistinguishable between the polar and non-polar molecules.

The positive $\Delta H$ indicates oil molecule desorption is endothermic, like oil droplet. But the $\Delta H$ is ~40% higher for the single oil molecules than the oil droplets per molecule. This is largely due to a fraction of oil molecules in the droplet do not interact with the kerogen surface directly, leading to lower $\Delta H$ value per molecule. This result suggests that the desorption free energy of a single molecule cannot be linearly scaled to a cluster of molecules or a droplet. The reduction of desorption enthalpy of oil droplets indicates that oil droplets are easier to desorb than single oil molecules.

As shown in Figure 3, the free energy of single oil molecule is less sensitive to the temperature with small $\Delta S$ values (7-8 J/mol/K) than that of the droplets (23-26 J/mol/K) (Figure 2 and Table 2). The calculated $\Delta S$ values for single oil molecules are about one R (gas constant), whereas, the $\Delta S$ for oil droplets is about ~3R. The significant difference highlights a dramatic gain in entropy of the oil droplets after desorbed from kerogen surface, which is largely attributed to the increased degree of freedom of the oil molecules. This result suggests that the desorption of a single oil molecule cannot be used to understand the desorption of an oil droplet and interfacial properties of oil-kerogen interaction in general. In addition, the result emphasizes that the entropic contribution to the free energy is significant for the desorption of oil droplets, which confirms the hypothesis of this study.

**Effect of Polarity of Oil Molecules**

Overall, the molecular polarity makes no significant differences in desorption free energy, entropy, and enthalpy except the desorption entropy of oil droplets (26.3 (3.2) J/mol/K for octane vs. 22.7 (1.4) J/mol/K for octanethiol). This difference is related to the less directional interaction (less restriction in translational and rotational mobility) in more hydrophobic nature of octane molecule than dipole-dipole interaction in octanethiol molecule. The slight differences in the desorption enthalpy are originated from the heterogeneous nature of kerogen surface containing both polar (hydroxyl -OH and amine -NH) and non-polar functional groups and presence of water molecules (hydrophilic) that compete with the oil molecules to interact with kerogen surface. As a result, the interactions of octane and octanethiol molecules with the surface are similar. Our previous simulations of the adsorption of single oil molecule on wet and dry kerogen surfaces showed that the surface water clearly alters the energetics of oil-water interactions.[24] On dry kerogen, a polar oil molecule requires nearly two times higher free energy to desorb for kerogen surface than that of non-polar one. On wet kerogen, water molecules take part at the surface interactions, suppressing interaction energetics of both polar and non-polar oil.[24] The result indicates the fluid composition can complicate the oil-surface interactions. This complexity suggests that the interactions between different components needs to be accurately described in molecular models in order to reliably simulate more realistic hydrocarbon systems with multiple chemical components.

**Contact Angle, Free Energy, and Surface Tension**

The contact angles of the polar and non-polar oil clusters on the kerogen surfaces are plotted in Figure 4. All the calculated contact angles are within 90°. Therefore, the kerogen surface can be considered as oil-wet at the temperatures from 300 K to 500 K. As temperature increases, the contact angles increase as well (Figure 4). This pattern is consistent with the decrease of desorption free energy as a function of temperature as shown in Figure 2, a result of positive entropy of desorption of the droplets from kerogen surface (Table 2). The calculated contact angle of the polar oil droplet is systematically smaller than those of non-polar oil by ~4° in the temperature range, from 48° to 64° for polar and 50° to 74° for non-polar oil. This difference suggests that the polar oil droplet is slightly more wettable by the kerogen surface than the non-polar oil droplet. Special attentions have been paid to make sure that the configurations used to calculate contact angles are corresponded to the free energy minimum of the system.[60] The reported contact angles should be free from the complications often caused by the contact angle hysteresis in experimental observations, i.e., the difference between advancing and receding contact angles due to the presence of metastable states.[60,61]

Contact angle is related to the surface tension ($\gamma$), which is the surface free energy ($F$) per surface area ($a$).[62–64]

$$\Gamma_{i\text{-}j} = (\partial F / \partial a_{i\text{-}j}) \tag{2}$$

$$dF = \gamma_{i\text{-}j} \cdot da_{i\text{-}j} \tag{3}$$

where $i$ and $j$ denote the phases in the simulations including water, kerogen surface, and oil.

It is reasonable to assume that the surface areas of kerogen remain constant since all the umbrella sampling simulations were performed at the equilibrium state.

Given that the contact angle ($\theta$) of oil droplet at kerogen surface can be described by the Young's equation.[65,66]

$$\gamma_{\text{water-kerogen}} - \gamma_{\text{oil-kerogen}} = \gamma_{\text{oil-water}} \cdot cos\theta \tag{4}$$

The desorption energy ($\Delta F$) can be expressed as

$$\Delta F = \gamma_{\text{oil-water}} \cdot (a_{\text{oil-water, free}} - a_{\text{oil-water, adsorbed}} + a_{\text{oil-kerogen, adsorbed}} \cdot cos\theta) \tag{5}$$

A detailed description about the derivation of equation (5) is provided in supporting text 3. At a given temperature, using the calculated free energy and interfacial areas from the MD simulation, equation (5) can be used to estimate surface tension of oil-water interface. However, estimating the surface area of the molecular cluster is not straight forward. Since the contact angle is directly measured based on the geometry of the molecular cluster adsorbed at the surface, it is reasonable to use geometric surface area that can be calculated in the similar way in the calculation of the contact angle. Geometric area indeed has been used in molecular modeling of surface wettability.[67 64] On the other hand, due to the relatively small number of molecules (~800 atoms) of the clusters, the geometry of oil droplets in this study is not smooth, especially at elevated temperatures. The increasing temperature will lead to noticeable changes of the surfaces as a result of the geometric deviation from an ideal spherical shape. Therefore, the surface areas were also estimated based on solvent-accessible surface area (SASA). The resulting calculated surface tensions between water and ~2 nm droplet are plotted in Figure 5, with higher numbers corresponding to calculations based on SASA values and lower ones corresponding to geometric surface area. It is well known that surface tension is dependent on the radius of the droplet.[68–71] As the radius of droplets falls below 30-50 nm and a sharp drop below 10 nm,[69–71] the calculated values of surface tension show a major drop from the value of flat interface. Estimated surface tension of n-$C_8H_{18}$ is reduced by ~50% for a ~2 nm droplet from the bulk value. Considering the size effect, the surface tensions calculated from the MD simulations are in good agreement with the experimentally measured surface tension between octane and water (Figure 5). In addition, the trend of surface tension as function of temperature is also reasonably predicted from the MD simulations, which shows a similar slope of the experimental data (Figure 5).

## Summary and Concluding Remarks

Molecular dynamics simulations were applied to investigate the temperature effect on the interactions of light oil with water-wetted kerogen. The free energy, enthalpy, and entropy of desorption as well as the contact angle of oil droplets were calculated. The result show that the free energy of oil/kerogen interactions is a linear function of temperature at shale reservoir temperatures. This relationship provides a molecular thermodynamic basis on the temperature effect on hydrocarbon reserves and in developing thermal stimulation techniques for unconventional shale. The simulations show that single molecules cannot represent an oil droplet based on a comparison of the free energies of desorption of droplets with those of single molecules because of the absence of oil-oil molecular interactions of the latter. This finding suggests that the system scale (i.e. pore size, slip length) is a critical variable in understanding the thermodynamics and making accurate predictions in nanoscales. The driver for the temperature effect on the desorption free energy is the entropy of desorption of oil molecules from kerogen surface. Because the entropy of surface desorption is closely related to the molecular gas-phase entropy, exploring the gas-phase entropy of oil molecules may provide important clues on heat induced enhanced oil recovery. Reasonable predictions of oil-water interfacial tension and its temperature dependence show the molecular models and molecular dynamics simulations are reliable for modeling of hydrocarbon-bearing fluid interactions with reservoir rock materials, and capable of filling the knowledge gap between experimental observations and theoretical calculations.

## Acknowledgment

The authors would like to thank Dr. Dipta Ghosh from Louisiana State University for meaningful discussions. This research used resources of the National Energy Research Scientific Computing Center (NERSC), a U.S. Department of Energy Office of Science User Facility operated under Contract No. DE-AC02-05CH11231. Portions of this research were conducted with high performance computational resources provided by the Louisiana Optical Network Infrastructure (http://www.loni.org).

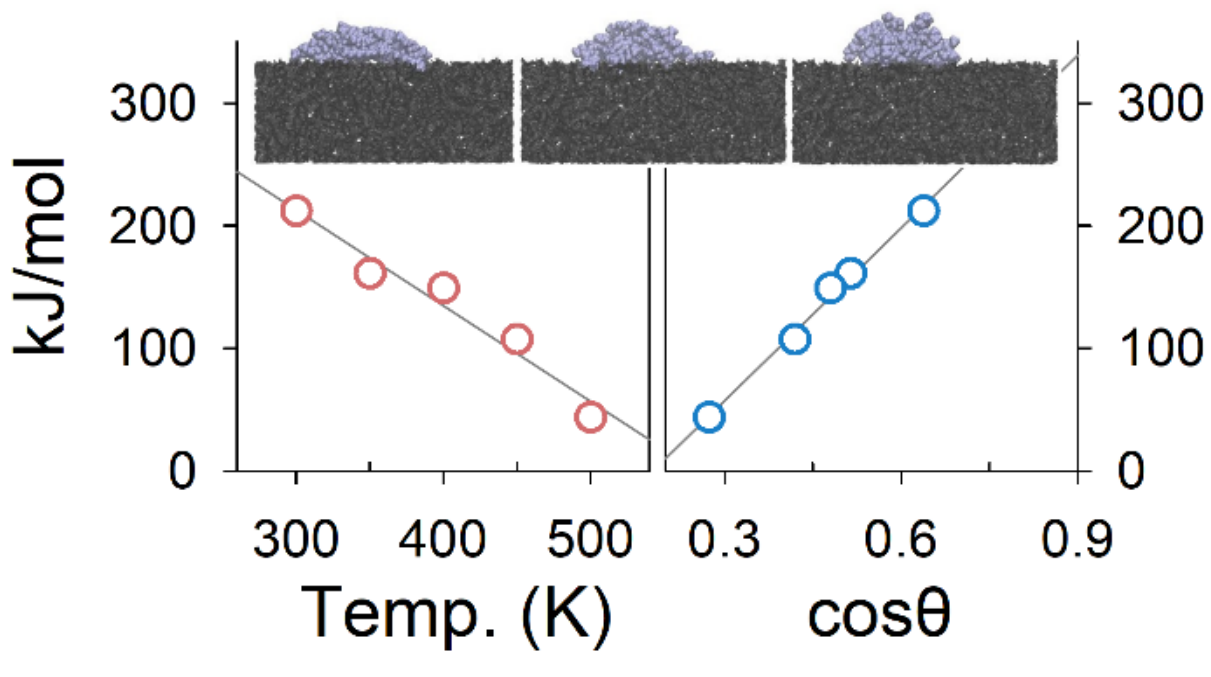

TOC abstract

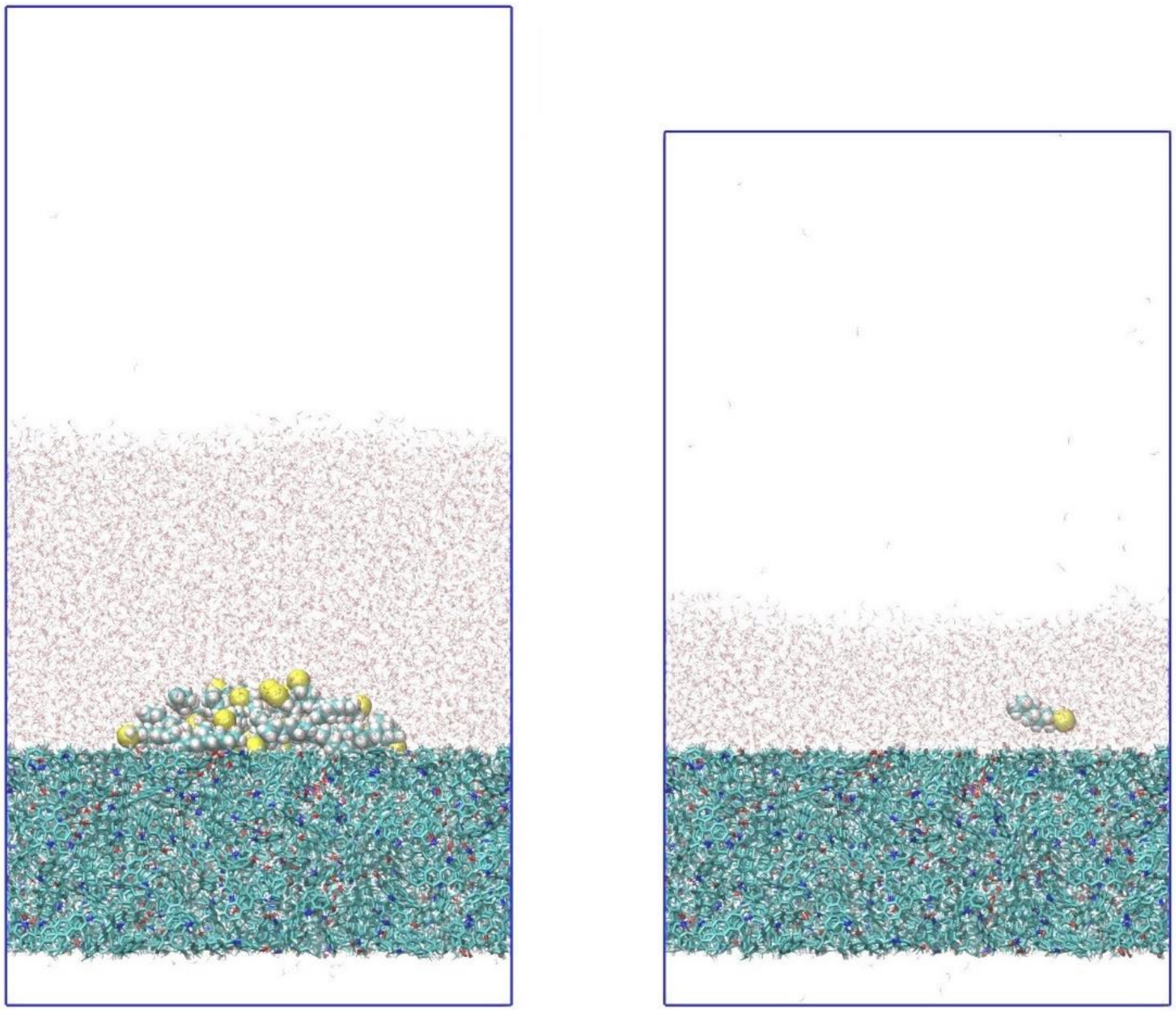

Figure 1. Simulation snapshot of a polar oil droplet (left) and a single polar oil molecule (right) on water-wetted kerogen surfaces. Each simulation box contains water, oil, and kerogen molecules. Different types of molecules are depicted with different styles for visual clarity. White represents hydrogen; green, carbon; yellow, sulfur; blue nitrogen; and red, oxygen.

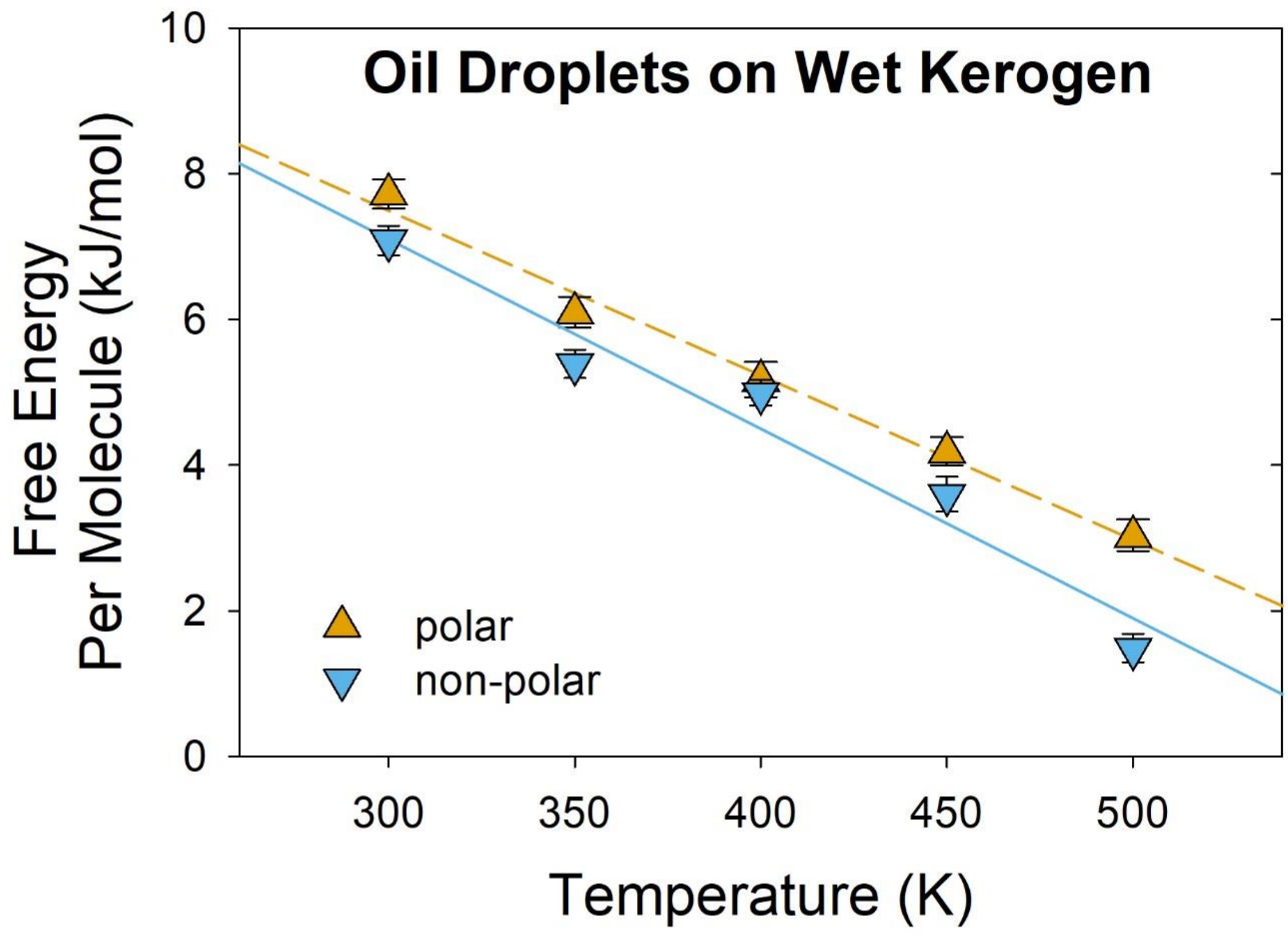

Figure 2. Temperature effect on the free energy of oil droplets on kerogen surfaces. The orange triangles (pointing upwards) represent free energy per molecule of a polar oil droplet, while the blue triangles (pointing downwards) denote the free energy of non-polar droplet. Standard errors are illustrated with error bars.

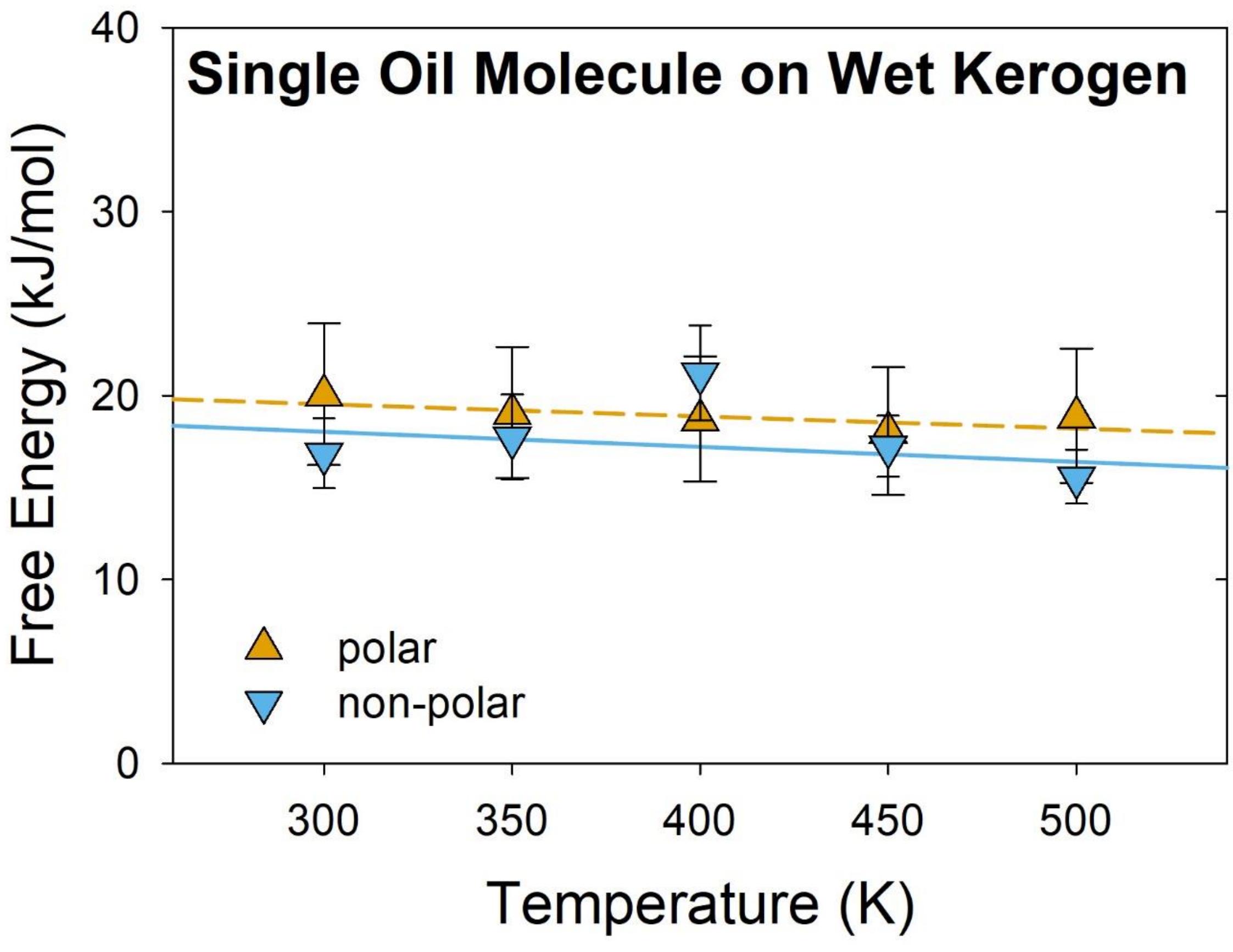

Figure 3. Temperature effect on the free energy of single oil molecules on kerogen surfaces. The orange triangles (pointing upwards) represent the polar oil, while the blue triangles (pointing downwards) denote the non-polar oil. Standard errors are illustrated with error bars.

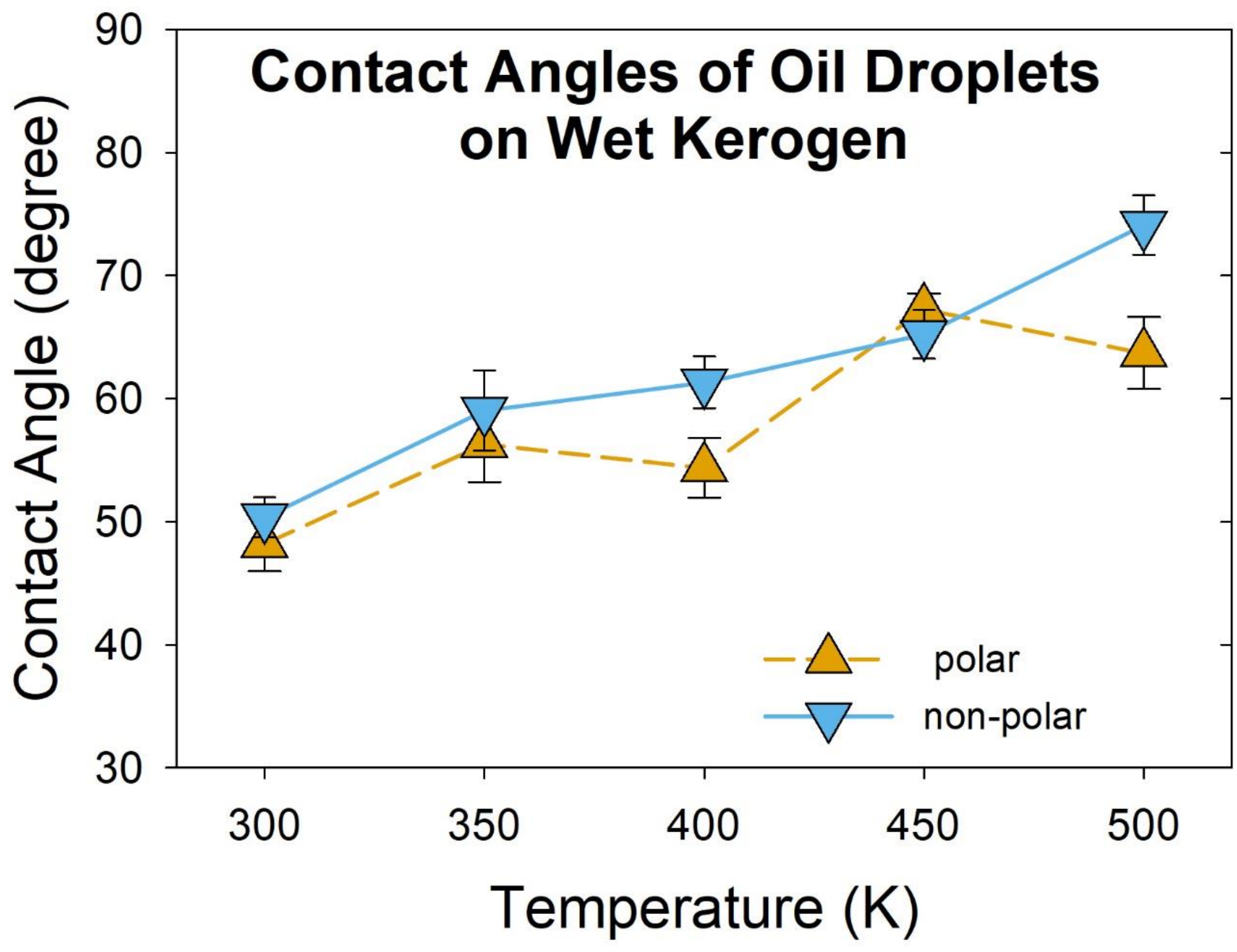

Figure 4. Temperature effect on the contact angle of polar and non-polar oil droplet on wet kerogen surfaces. The orange triangles (pointing upwards) represent the polar oil, while the blue triangles (pointing downwards) denote non-polar. Standard errors are illustrated with error bars.

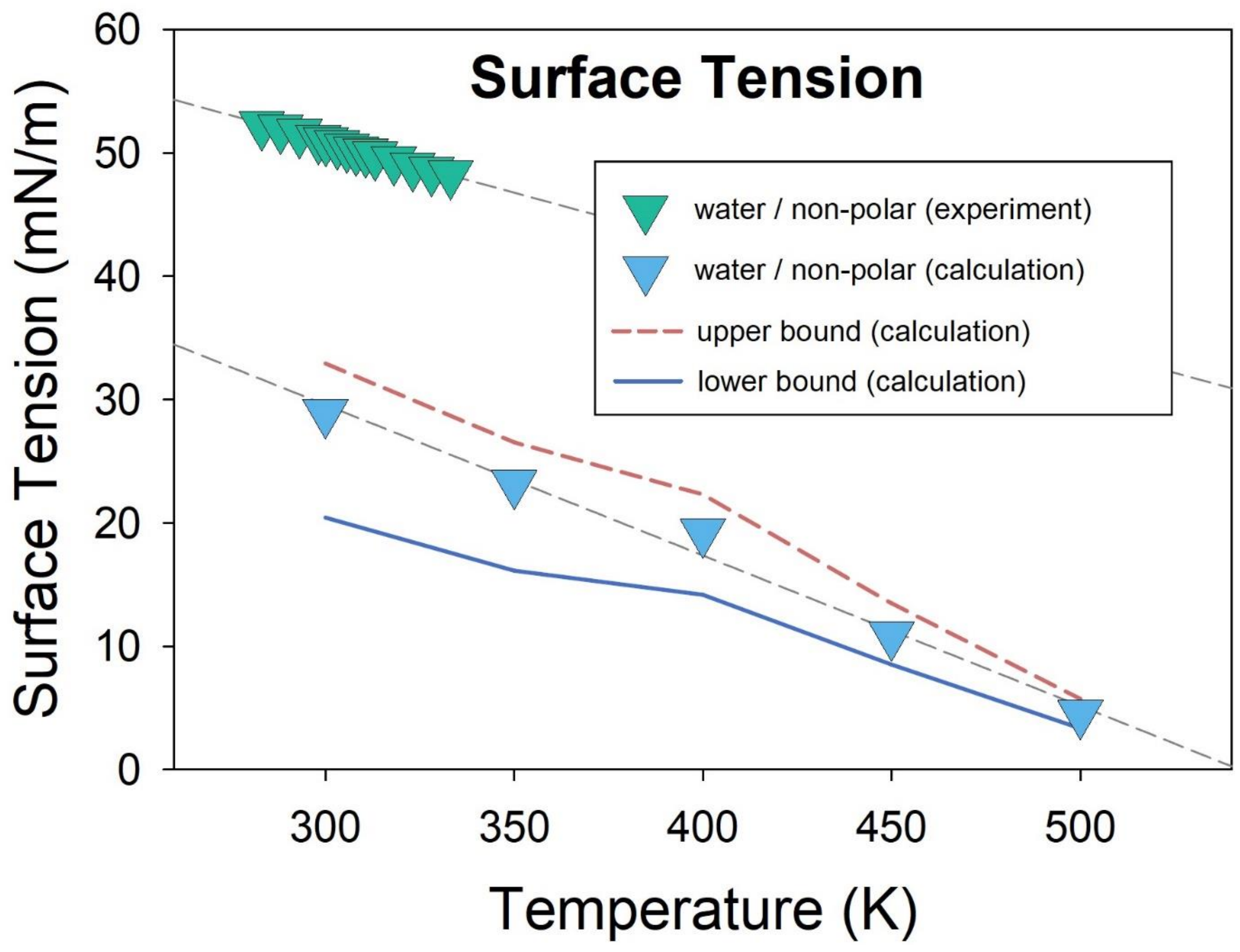

Figure 5. Calculated surface tension of water-oil with respect to temperature. The orange triangles (pointing upwards) represent the surface tension of water/polar oil, the blue triangles (pointing downwards) denote surface tension of water/non-polar oil, the hollow blue triangles (pointing downwards) stand for the experimental measurements at atmospheric pressure.[72]

| System | Single oil molecule | Oil droplet ( 30 molecules) |
|---|---|---|
| Box dimension (X × Y × Z, nm) | 8.1 × 7.9 × 14.0 | 8.1 × 7.9 × 16.0 |
| Water molecules | 3950 | 10000 |
| Slab thickness (nm) | 3.54 | 3.54 |
| Total number of atoms | $3.0 \times 10^5$ | $5.0 \times 10^5$ |
| Configurations | 261 | 361 |
| Spacing (nm) | 0.01 | 0.01 |
| Harmonic potential (kJ/mol / $nm^2$) | 5000 | 5000 |
| Production time (ns) | 0.2 | 0.1 |

Table 1. Specification of simulation systems, which includes simulation box size, number of water molecules, kerogen slab thickness, total number of atoms, number of configurations for umbrella sampling, the spacing between windows, harmonic potential, and production time of each configuration.

| System | ΔG (kJ/mol) | | | | | ΔH | ΔS |
|---|---|---|---|---|---|---|---|
| | 300K | 350K | 400K | 450K | 500K | kJ/mol | J/mol/k |
| Polar molecule | 20.1 (3.9) | 19.1 (3.6) | 18.7 (3.4) | 18.1 (3.4) | 18.9 (3.7) | 21.6 (1.5) | 6.6 (3.7) |
| Non-polar molecule | 16.9 (1.9) | 17.7 (2.3) | 21.2 (2.6) | 17.2 (1.7) | 15.6 (1.5) | 20.6 (4.8) | 8.2 (11.4) |
| Polar droplet * | 7.73 (0.20) | 6.10 (0.21) | 5.17 (0.24) | 4.19 (0.20) | 3.04 (0.22) | 14.3 (0.5) | 22.7 (1.4) |
| Non-polar droplet * | 7.09 (0.20) | 5.39 (0.20) | 4.99 (0.18) | 3.60 (0.24) | 1.49 (0.19) | 15.0 (1.3) | 26.3 (3.2) |

Table 2. Changes of free energy (ΔG), enthalpy (ΔH), and entropy (ΔS) calculated from the oil interactions with wet kerogen surfaces under different temperatures. * The ΔG, ΔH, and ΔS of oil droplets are normalized as per molecule for comparison. Their original values, based on the whole oil cluster, are divided by 30, the total number of molecules in the oil clusters.

# Temperature Effect on the Interactions between Oil Droplet and Kerogen Surface at Reservoir Temperatures

Zelong Zhang,*,† Adrienne Stephens,† and Jianwei Wang†,§

†Department of Geology and Geophysics, Louisiana State University, Baton Rouge, LA 70803, United States

§Center for Computation and Technology, Louisiana State University, Baton Rouge, LA 70803, United States

Corresponding to: zhangzelong@protonmail.com

List of Contents

**Supporting Text 1. Parameterization of SPC/F water model.**

```
[ moleculetype ]
; molname      nrexcl
SOL            2

[ atoms ]
;   nr     type  resnr residue  atom   cgnr     charge       mass
     1       OW      1    SOL     OW      1     -0.820      15.99940
     2       HW      1    SOL    HW1      1      0.410       1.00800
     3       HW      1    SOL    HW2      1      0.410       1.00800

[ bonds ]
; i     j      funct   length  force.c.
1       2       1       0.09572 502080.0 ; OW HW
1       3       1       0.09572 502080.0 ; OW HW

[ angles ]
; i     j       k       funct   angle   force.c.
2       1       3       1       109.50  627.600 ; HW-OW-HW
```

**Supporting Text 2. Tcl/tk script for VMD**

Scripts and instruction are available at https://github.com/er1czz/vmd

**Supporting Text 3. Equation derivation.**

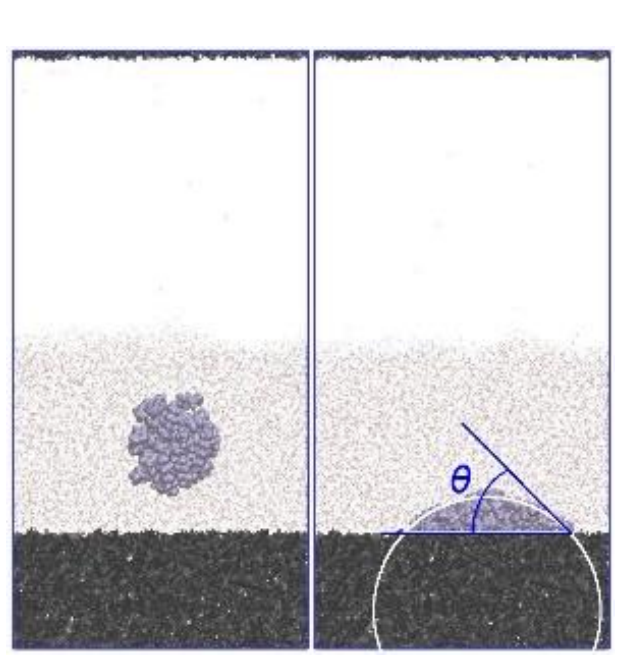

Free energy ($F$) can be expressed as a function of surface tension ($\gamma$) and surface area ($a$), which can be applied for both the Helmholtz and the Gibbs free energy. In the equation below, “$_{o}$” denotes oil; “$_{s}$”, surface; “$_{w}$”, water; and “$_{v}$”, vacuum. And “ $^{o}$ ” and “ $'$ ”stand for a system with the free oil droplet and a system with the adsorbed oil droplet, respectively.

For both systems, $dF = \gamma_{\text{i-j}} \cdot da_{\text{i-j}}$ can be expanded as

$$dF = \gamma_{\text{s-v}} \cdot da_{\text{s-v}} + \gamma_{\text{w-v}} \cdot da_{\text{w-v}} + \gamma_{\text{o-w}} \cdot da_{\text{o-w}} + \gamma_{\text{s-w}} \cdot da_{\text{s-w}} + \gamma_{\text{o-s}} \cdot da_{\text{o-s}} \qquad \text{(s1)}$$

For systems in steady/equilibrium state, the surface areas of water-vacuum and surface-vacuum contact regions should be unchanged. Therefore

$$0 = da_{\text{w-v}} = da_{\text{s-v}} \qquad \text{(s2)}$$

Given that the free energy of desorption

$$\Delta F = F^{o} - F' \qquad \text{(s3)}$$

Combine s1, s2, and s3 to obtain

$$\Delta F = (\gamma_{\text{o-w}} \cdot a^{o}_{\text{o-w}} + \gamma_{\text{s-w}} \cdot a^{o}_{\text{s-w}}) - (\gamma_{\text{o-w}} \cdot a'_{\text{o-w}} + \gamma_{\text{s-w}} \cdot a'_{\text{s-w}} + \gamma_{\text{o-s}} \cdot a'_{\text{o-s}}) \qquad \text{(s4)}$$

Since kerogen surface is rigid, its total surface area should be a constant value

$$a^{o}_{\text{s-w}} = a'_{\text{s-w}} + a'_{\text{o-s}} \qquad \text{(s5)}$$

Substituting for $a^{o}_{\text{s-w}}$ in s4 gives

$$\Delta F = \gamma_{\text{o-w}} \cdot (a^{o}_{\text{o-w}} - a'_{\text{o-w}}) + (\gamma_{\text{s-w}} - \gamma_{\text{o-s}}) \cdot a'_{\text{o-s}} \qquad \text{(s6)}$$

According to the Young’s equation

$$\gamma_{\text{s-w}} - \gamma_{\text{o-s}} = \gamma_{\text{o-w}} \cdot cos\theta \qquad \text{(s7)}$$

Now substitute for $\gamma_{\text{s-w}} - \gamma_{\text{o-s}}$ in s6 we obtain

$$\Delta F = \gamma_{\text{o-w}} \cdot (a^{o}_{\text{o-w}} - a'_{\text{o-w}} + a'_{\text{o-s}} \cdot cos\theta) \quad \text{QED.} \qquad \text{(s8)}$$

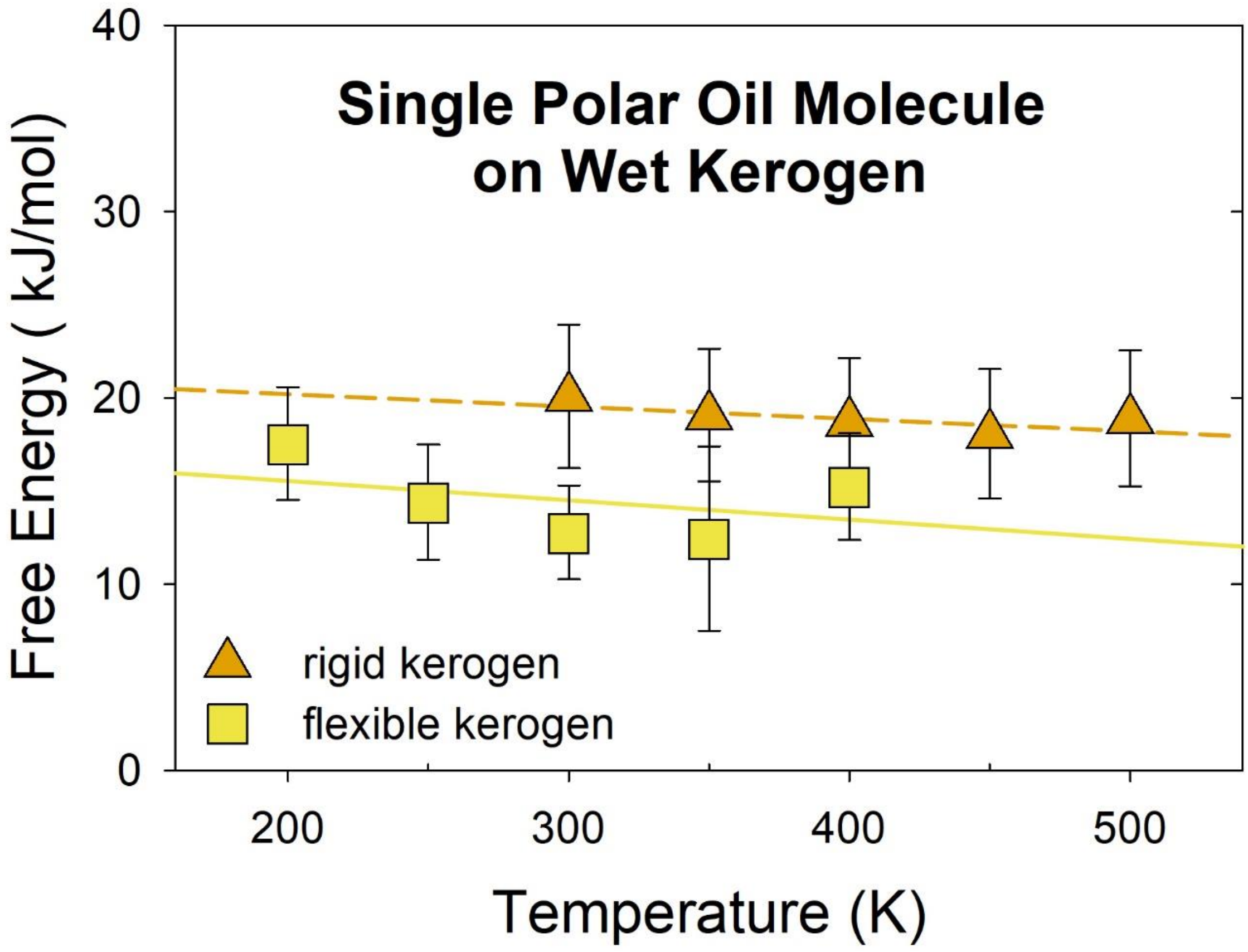

**Figure S1.** Comparison between the free energy of a single polar oil molecule desorption from a fixed (rigid) kerogen surface and from an unrestricted (flexible) kerogen surface under different temperatures. The energies of oil desorption from unrestricted kerogen surfaces are systematically lower than those from fixed kerogen surfaces. Standard errors are illustrated with error bars.

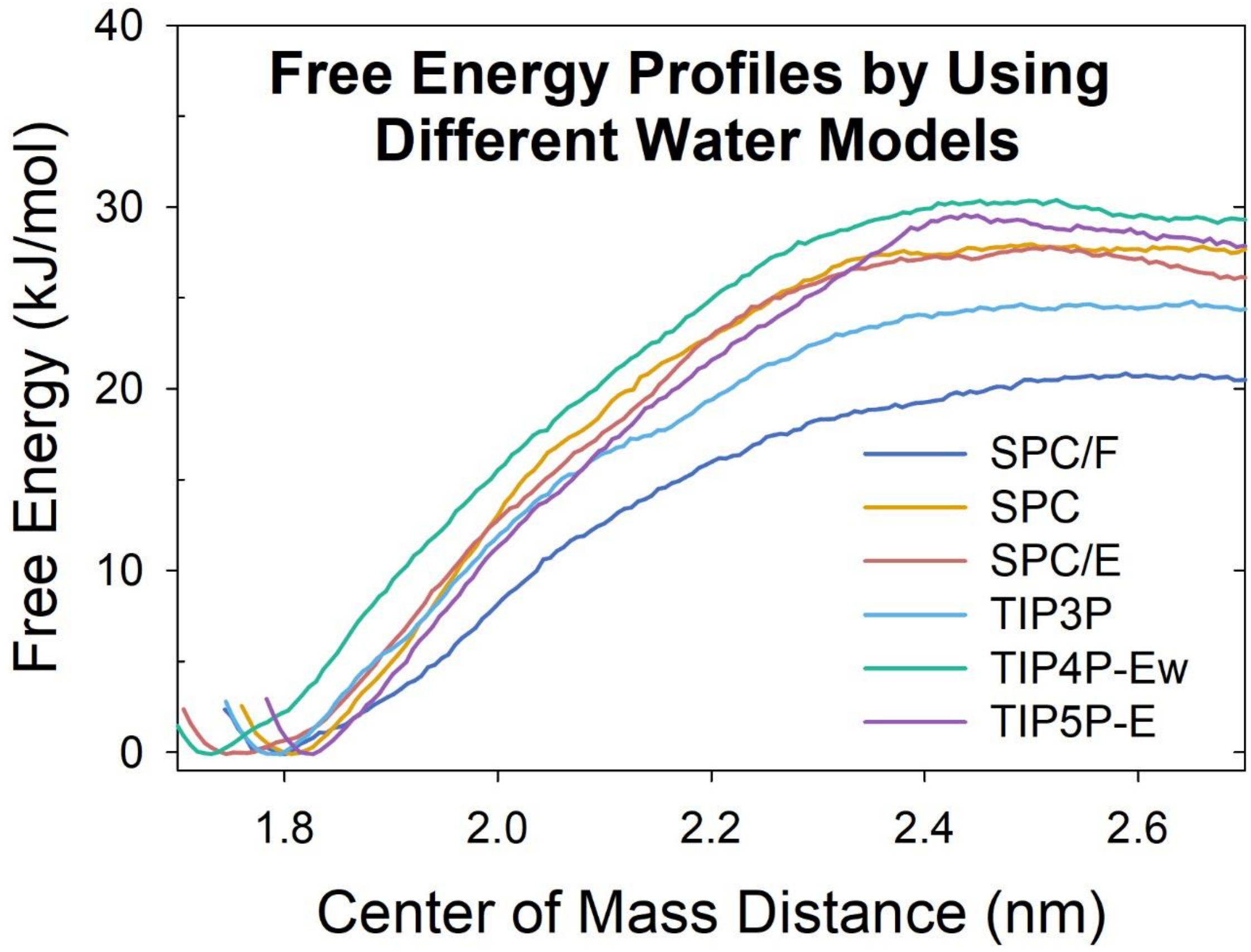

**Figure S2.** The free energy profiles of a single polar oil molecule on a rigid kerogen surfaces wetted with different water models. The distance is measured from the center of mass of the oil molecule to that of the kerogen slab. These free energy profiles are qualitatively identical. The deviation in energy minimum position is reasonable, and is attributed to the heterogeneity of kerogen surfaces and the asymmetric geometry of the polar oil molecule.

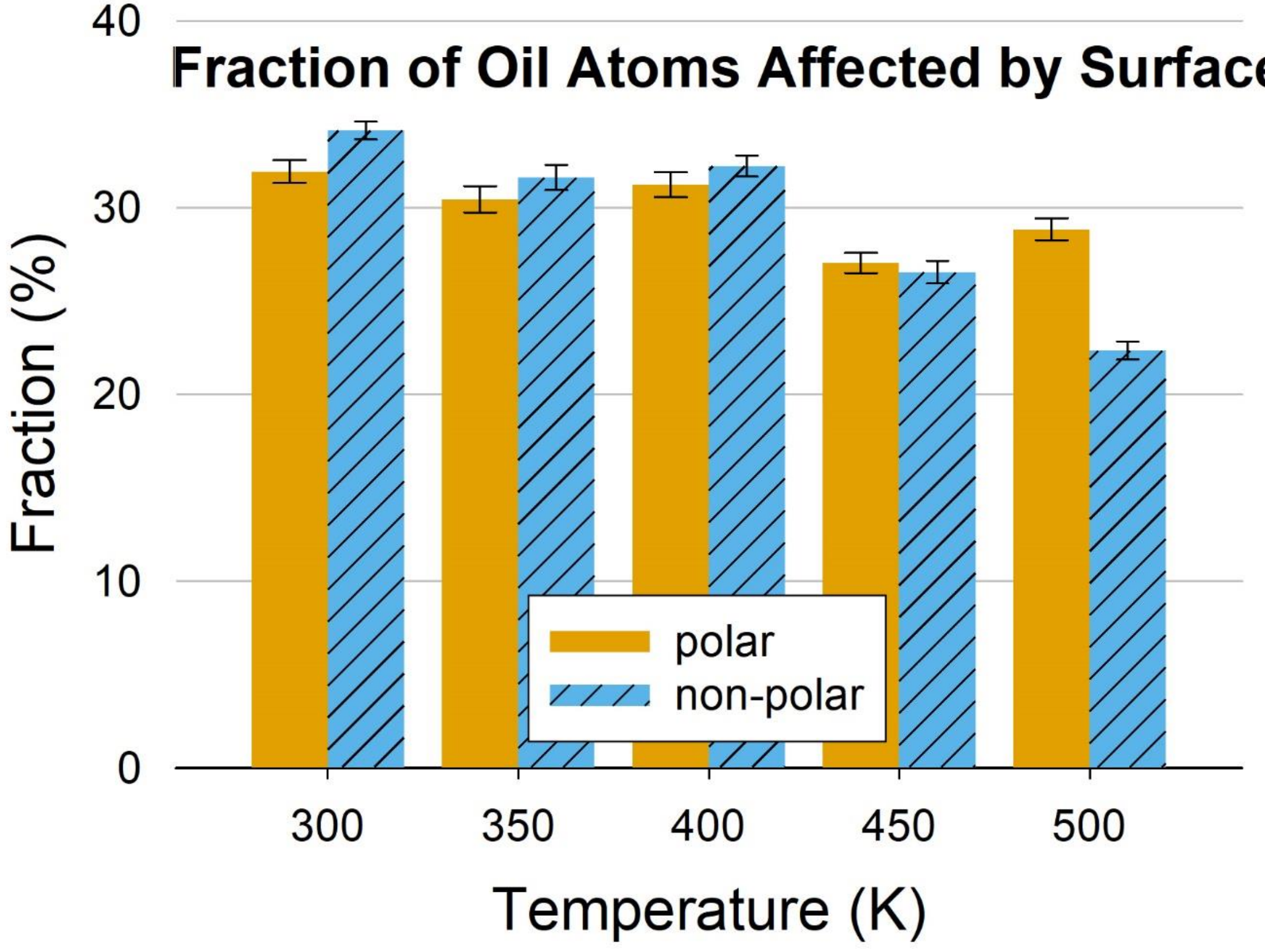

**Figure S3.** Fraction of number of atoms in oil droplets affected by the kerogen surface. The fraction is the percentage of oil atoms within 0.36 nm of the surface out of the total number of atoms in the oil droplet. The orange bars represent polar oil droplets, while the blue bars (with stripe pattern) denote the non-polar oil droplets. Standard errors are illustrated with error bars.

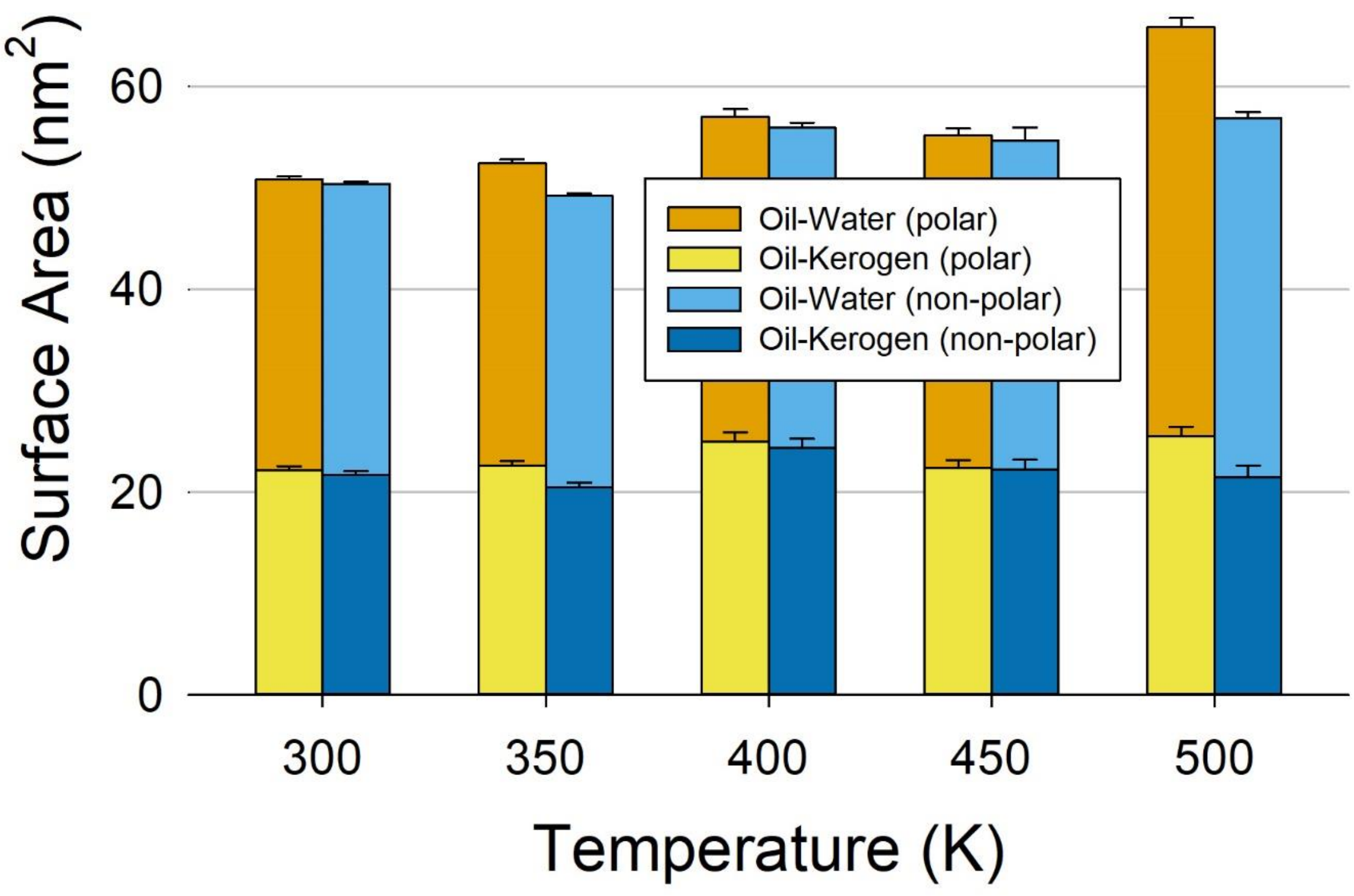

**Figure S4.** Surface areas of different contact regions on adsorbed oil droplets. Orange, yellow, sky blue, and blue squares denote contact areas between polar oil and water, polar oil and kerogen surface, non-polar oil and water, and non-polar oil and kerogen surface, respectively. Standard errors are illustrated with error bars.

| Model | Description | Dipole | Hours | Efficiency % | ΔG (kJ / mol) |
|---|---|---|---|---|---|
| SPC/F | Flexible Simple Charge Point, 3 sites | 2.39 – 2.46[1,2] | 48 | 100 | 19.08 (3.57) |
| SPC | Simple Charge Point, 3 sites | 2.27[3] | 49 | 98 | 27.99 (3.36) |
| SPC/E | Extended Simple Charge Point, 3 sites | 2.35[4] | 49 | 98 | 27.85 (2.70) |
| TIP3P | Transferable Intermolecular Potential, 3 sites | 2.35[5] | 49 | 98 | 24.85 (2.61) |
| TIP4P-Ew | 4 sites, Ewald summation | 2.32[6] | 54 | 89 | 30.41 (3.21) |
| TIP5P-E | 5 sites, Ewald summation | 2.29[7] | 62.5 | 77 | 29.60 (3.00) |

**Table S1.** Description and performance of water models SPC/F, SPC, SPC/E, TIP3P, TIP4P-Ew, and TIP5P-E. The benchmark tests were conducted at NERSC using 64 cores on Cori Intel Xeon Phi (KNL) node. Each calculation computed a free energy surface at 350 K of a single polar oil molecule interacting with a fixed kerogen surface in the presence of 3950 water molecules. Each free energy surface was derived from 121 configurations. 200 ps umbrella sampling was performed on each configuration.

| Compound | Dipole Moment (Debye) | | |
|---|---|---|---|
| | MD | PM7 | DFT |
| water | 1.13 | 2.130 | 2.094514 |
| octane | 0.00 | 0.003 | 0.000161 |
| octanethiol | 1.05 | 2.218 | 2.025873 |
| kerogen | 1.27 | 0.490 | 0.645501 |

**Table S2.** Molecular dipole moment of water, octane, octanethiol, and kerogen. Molecular dynamics simulation (MD) were carried out by Avogadro.[8,9] Semi-empirical calculations were conducted by MOPAC with PM7 Hamiltonian.[10,11] Density function theory (DFT) calculations were performed by GAMESS with 6-31G* basis set,[12,13] B3LYP functional,[14,15] and a SCF convergence of $10^{-5}$ of the density matrix. Please note that all these calculations are based on single molecule models in vacuum. The optimized kerogen structure can be downloaded from https://github.com/er1czz/md

| Temperature (K) | Contact angle (degree) | |
|---|---|---|
| | Polar oil | Non-polar oil |
| 300 | 48.27 (2.26) | 50.37 (1.62) |
| 350 | 56.35 (3.12) | 59.05 (3.27) |
| 400 | 54.38 (2.42) | 61.33 (2.10) |
| 450 | 67.29 (1.27) | 65.24 (1.98) |
| 500 | 63.72 (2.92) | 74.12 (2.42) |

**Table S3.** Contact angles of polar and non-polar oil droplets at different temperatures. "()" denotes standard error.

| Temperature (K) | Polar oil droplet system pressure (bar) | | Non-polar oil droplet system pressure (bar) | |
|---|---|---|---|---|
| | Free | Adsorbed | Free | Adsorbed |
| 300 | 2.00E+03 | 1.98E+03 | 2.03E+03 | 2.01E+03 |
| 350 | 2.08E+03 | 2.04E+03 | 2.14E+03 | 2.13E+03 |
| 400 | 2.18E+03 | 2.15E+03 | 2.21E+03 | 2.19E+03 |
| 450 | 2.32E+03 | 2.28E+03 | 2.34E+03 | 2.31E+03 |
| 500 | 2.49E+03 | 2.45E+03 | 2.51E+03 | 2.48E+03 |

**Table S4.** System average pressures of free oil droplet simulations and adsorbed oil droplet simulations at different temperatures.

| Temperature | Polar oil ($nm^2$) | | | Non-polar oil ($nm^2$) | | |
|---|---|---|---|---|---|---|
| | Free droplet | Adsorbed droplet | | Free droplet | Adsorbed droplet | |
| | Oil-water | Oil-water | Oil-kerogen | Oil-water | Oil-water | Oil-kerogen |
| 300 K | 36.27 (0.44) | 28.66 (0.30) | 22.18 (0.37) | 34.84 (0.49) | 28.71 (0.18) | 21.71 (0.35) |
| 350 K | 37.66 (1.14) | 29.86 (0.37) | 22.59 (0.48) | 35.66 (0.35) | 28.76 (0.24) | 20.49 (0.43) |
| 400 K | 43.39 (1.02) | 32.06 (0.73) | 24.98 (0.90) | 40.14 (1.00) | 31.63 (0.49) | 24.32 (0.96) |
| 450 K | 45.48 (1.02) | 32.86 (0.64) | 22.36 (0.78) | 45.24 (1.22) | 32.51 (1.25) | 22.20 (0.99) |
| 500 K | 55.77 (1.06) | 40.39 (0.91) | 25.48 (0.95) | 49.42 (1.59) | 35.40 (0.66) | 21.46 (1.16) |

**Table S5**. Surface areas of different contact regions on oil droplets. “()” denotes standard error.